\documentstyle{article}
\begin{document}

\author{V.E. Rochev\\
{\it Institute for High Energy Physics,}\\ {\it 142284
Protvino,
Moscow region, Russia}}
\title{On Higgs Mechanism in Non-Perturbative Region}
\date{}
\maketitle
\begin{abstract}
Generalization of the Higgs mechanism  which takes
into account the contributions of
gauge field vacuum configuration into the formation
of the physical vacuum is considered.
 For the Abelian Higgs model the triviality bound
 $m_H\le~1.15m_A$ is found.\\

\end{abstract}

\section{Introduction}
Higgs mechanism is one of the crucial points of Standard
Model
and simultaneously  one of the most mysterious its
properties.
Considerable efforts on the experimental search for Higgs
particles have not still lead to  success\cite{LEP}.
Theoretical investigation of the scalar sector of the
Standard
Model is also far from  completeness. In attempting to
go beyond
the framework of the quasiclassical approximation and  the
perturbation theory, one encounters a number of
difficulties
and  the principal problem of them is the well-known
triviality of quadric
scalar self-interaction (see \cite{Callaway} and refs.
therein):
the renormalized coupling constant of
$\phi^4_4$-interaction tends
to zero at the cutoff removing.
The triviality of $\phi^4_4$-interaction leads to the fact,
that the mass of Higgs particle is not a fully independent
parameter but it is connected with other parameters
of the model
such as intermediate boson masses, $t$-quark mass, etc.
In the frameworks of different approaches (see, for example,
 \cite{Callaway}, \cite{Triv} and refs. therein) this
 fact leads
to  different estimates of the Higgs boson mass, and
the absence of experimental data does not
 favor over any approach.
The common feature for all investigations of the Higgs
mechanism
is the fundamental proposition about the existence of
non-zero
Higgs field vacuum expectation value
\begin{equation}
<0\mid\phi\mid 0> \equiv v \neq 0,
\end{equation}
which defines the Higgs boson and gauge field masses.
The existence of such a non-zero expectation value means
that
the physical vacuum of the electroweak interaction is
a non-trivial medium --- the relativistic superconductor.
The structure of this medium is not investigated in full
measure,
and there are no prior foundations to suppose that such a
relativistic medium
is completely similar to non-relativistic Ginzburg-Landau
superconductor.
For example, the triviality of the $\phi^4_4$-interaction
is
a purely relativistic quantum field phenomenon, which
has no
analogue in non-relativistic case.
Usually the role of the gauge field
in the formation of the physical vacuum is fully ignored
when
one considers the Higgs mechanism.
This is made on the base of a quite evident observation
that
the non-zero
vacuum expectation values of the vector gauge field destroy
the Poincar\'{e}-invariance of the theory. On the other
hand,
however, it is necessary to take into account that the
quantum
field theory equations possess the variety of solutions,
and the choice of a unique physical solution  is, in
fact, the
definition of the above-mentioned relativistic medium
--- the
physical vacuum. The example is the Higgs mechanism
itself
in its traditional formulation: besides the solution over
non-symmetrical physical vacuum (1), the usual symmetrical
solution over the trivial ("perturbative") vacuum
always exists,
and the choice  in favor of either solution  is defined
by the sign of the quadratic term coefficient of Higgs field
Lagrangian.
But the complete set of solutions of quantum field
theory equations
is not exhausted by these two classes of solutions.
A variety
of other solutions exists, and
among them there are those with non-zero expectation
values
of vector field.
Due to linearity of basic equations the general solution is
a superposition of the partial solutions including the
solutions
with non-zero expectation values of vector field. Each
of the
partial solutions defines its "partial mode". A candidate
for the physical
vacuum is a superposition of partial modes, which
satisfies
the certain physical conditions including the
Poincar\'{e}-invariance.
The Poincar\'{e}-invariance of the theory does
not mean, generally speaking, the absence of the
partial modes
with non-zero vacuum expectation of the gauge field in this
superposition
but only means the mutual cancellation of their
contributions  in the
 physical vacuum expectation value of the gauge field.
In the present work the generalization of the Higgs
mechanism is
considered, which takes into account the contribution of the
partial solutions of quantum field theory equations
(Schwinger-Dyson equations) with non-zero expectations
of both scalar
and vector fields. A method of iteration solution of
Schwinger-Dyson
equations for the generating functional of Green
functions, which
permits taking into account the contribution of
such solutions,
was developed in works \cite{Ro1}-\cite{Ro4}.
This method does not require the effective potential
be introduced,
whose use for a theory with vector fields run across  the
known difficulties.
The method has been applied to the description of
spontaneous
symmetry breaking for scalar theory \cite{Ro1}, \cite{Ro2}
and Gross-Neveu model at $N$ finite \cite{Ro2}, \cite{Ro3},
and also for non-Abelian gauge theory  \cite{Ro4}
(for constant vacuum configurations).
In Section 2 the method is described by an example of
 the complex scalar field. In section 3 the method is
applied to the Abelian Higgs model. Taking into account
the vacuum configurations of gauge fields allows one
to model
the  triviality of the $\phi^4_4$-interaction. In this case
a very strict triviality bound  $(m_H\le~1.15m_A)$
arises. In the concluding Section  a possibility of
generalization
for non-Abelian theory is briefly discussed.

\section{Complex scalar field with self-action}
As the first step to the Higgs model and for the
illustration
of the investigation method consider  the theory of the
complex scalar field $\phi$ with Lagrangian
\begin{equation}
{\cal L} = \partial_\mu \phi^* \partial_\mu \phi -
m^2 \phi^* \phi - \frac{\lambda}{2} (\phi^* \phi)^2.
\end{equation}
We work in Minkowski space with
$g_{\mu\nu} = \mbox{diag} (1,-1,-1,-1)$,
but for the notation simplicity do not distinguish
the upper and lower
Lorentz vector indices.
The generating functional of Green functions $G(j)$
is a solution of
Schwinger-Dyson equation (SDE)
\begin{equation}
\frac{\lambda}{i}\frac{\delta^3 G}{\delta j^*
\delta j \delta j^*}
- (m^2 + \partial^2) \frac{1}{i} \frac{\delta G}
{\delta j^*}
+ j G = 0.
\end{equation}
Here $j(x)$ is a source of field $\phi^* (x)$.
Green functions (vacuum expectation values of
$T$-product)
are the functional derivatives of $G$ at the
source switched off.
At $\lambda = 0$ equation (3) has a solution
$$
G^{pert} = \exp \{ i \int dx dy j^*(x)
\Delta_c(x-y) j(y)\},
$$
where $\Delta_c = (m^2 + \partial^2)^{-1}$ is the free
propagator.
This solution is the foundation for the
iteration scheme of the coupling constant perturbation
theory.
We shall use an iteration scheme of solving SDE which
is based on an alternative
principle and allows one to investigate the
non-perturbative effects
(see also \cite{Ro1} -- \cite{Ro4}).
This iteration scheme is based on the following
considerations:
since the purpose of the calculations is the Green
functions, it is
sufficient for us to know the generating functional
$G(j)$ near the
point  $j=0$. Therefore it is reasonable to take as a
leading
approximation of SDE an equation with "constant
coefficients", i.e.
to put $j=0$ in the coefficient functions of eq.(3).
The leading
approximation equation has the form
\begin{equation}
\frac{\lambda}{i}\frac{\delta^3 G_0}{\delta j^*
\delta j \delta j^*}
- (m^2 + \partial^2) \frac{1}{i}
\frac{\delta G_0}{\delta j^*} = 0.
\end{equation}
This equation has a solution
\begin{equation}
G_0 (j) = \exp i\int dx [v^*(x)j(x) + v(x)j^*(x)],
\end{equation}
where $v(x)$ obeys the "characteristic equation"
\begin{equation}
(\lambda v^* v + m^2 + \partial^2)v = 0.
\end{equation}
Functional (5), considered as a leading ("vacuum")
approximation,
is a foundation for the linear iteration scheme
$$
G = G_0 + G_1 + \cdots + G_n + \cdots,
$$
where $G_n$ is defined by the iteration scheme equation
\begin{equation}
\frac{\lambda}{i}\frac{\delta^3 G_n}{\delta j^*
\delta j \delta j^*}
- (m^2 + \partial^2) \frac{1}{i} \frac{\delta G_n}
{\delta j^*} =
- j G_{n-1} .
\end{equation}
A solution of eq.(7) is the functional
$$
G_n(j) = P_n(j) G_0(j),
$$
where $P_n(j)$, with taking into account the leading
approximation
equations (5)-(6), is a solution of the equation
\begin{eqnarray}
\frac{1}{i}(2\lambda v^* v + m^2 +
 \partial^2)\frac{\delta P_n}{\delta j^*}
+\frac{\lambda v^2}{i}\frac{\delta P_n}{\delta j}-\\
- \lambda (2v\frac{\delta}{\delta j} + v^*\frac{\delta}
{\delta j^*})
\frac{\delta P_n}{\delta j^*} -
\frac{\lambda}{i}\frac{\delta^3 P_n}
{\delta j^* \delta j \delta j^*}
= jP_{n-1}. \nonumber
\end{eqnarray}
Since $P_0 =1$, the solution of eq.(8)  is a polynomial
in the source $j$ at any $n$. Coefficient functions of this
polynomial
define the Green functions of the corresponding step
of the
iteration scheme. Thus, the solution of the
first-step equation
is a quadratic polynomial in $j$:
$$
P_1(j) =
i\int dx dy \{j^*(x)\Delta(x,y)j(y)
 + \frac{1}{2}
\bigl(j(x)d(x,y)j(y)+
$$
$$
 + j^*(x)\bar d(x,y)j^*(y)\bigr)\}
+i\int dx \bigl(\bar \Phi(x)j(x) + j^*(x)\Phi(x)\bigr).
$$
The solution of the second-step equation is a polynomial
of the fourth order and
so on. At any step equation (8) defines  a closed system of
equations for coefficient functions of the
polynomial $P_n(j)$.
For the ultraviolet divergences removing it
is necessary to modify
SDE by introducing  counterterms,
that is by making the following substitution in eq.(3):
$$
\lambda\rightarrow\lambda+\delta\lambda,\;m^2\rightarrow
m^2+\delta m^2,\;
\partial^2\rightarrow (1+\delta z)\partial^2.
$$
Here $\delta\lambda, \delta m^2$ and $\delta z$
are counterterms of
the coupling constant, mass and wave function
renormalization.
Eq.(7) and eq.(8) for $P_n$ are modified by
introducing  counterterms of corresponding orders of the
iteration scheme.
There is no need to introduce the counterterms
in the leading order,
and leading approximation equation (4) and l.h.s.
of eq.(7) are
not changed. The first-step equation with the counterterms
has the form
$$
\frac{\lambda}{i}\frac{\delta^3 G_1}{\delta j^*
\delta j \delta j^*}
- (m^2 + \partial^2) \frac{1}{i} \frac{\delta G_1}
{\delta j^*}=
$$
$$
= - j G_0 -
\frac{\delta\lambda_1}{i}\frac{\delta^3 G_0}{\delta j^*
\delta j \delta j^*}
+ (\delta m^2_1 + \delta z_1\partial^2)
\frac{1}{i} \frac{\delta G_0}{\delta j^*},
$$
and, correspondingly, the equation for $P_1$ is modified as
$$
\frac{1}{i}(2\lambda v^* v + m^2 +
\partial^2)\frac{\delta
P_1}{\delta j^*}
+\frac{\lambda v^2}{i}\frac{\delta P_1}{\delta j}-
 \lambda(2v\frac{\delta}{\delta j} +
 v^*\frac{\delta}{\delta j^*})
\frac{\delta P_1}{\delta j^*}=
$$
$$
= j - \delta\lambda_1 v^*v^2 -
(\delta m^2_1 + \delta z_1
\partial^2)v.
$$
For the first-step coefficient functions
$\Delta,\, d$ ³ $\Phi$
we obtain the system of equations
\begin{equation}
(2\lambda v^* v + m^2 + \partial^2)\Delta (x,y) +
\lambda v^2 d(x,y)
= \delta (x-y),
\end{equation}
\begin{equation}
(2\lambda v^* v + m^2 + \partial^2)d(x,y) + \lambda (v^*)^2
\Delta (x,y) = 0,
\end{equation}
\begin{equation}
(2\lambda v^* v + m^2 + \partial^2)\Phi +
\lambda v^2 \bar \Phi -
i\lambda(2v\Delta (x,x) + v^* \bar d(x,x)) =
\end{equation}
$$
= -\delta\lambda_1 v^*v^2 - (\delta m^2_1 +
\delta z_1 \partial^2)v.
$$
The particle propagators are defined by subsystem (9)-(10),
while eq.(11) is a connection for first-step counterterms.
Note, that solutions of eqs.(9) and (10) are evidently
ultraviolet-finite,
and therefore the first-step counterterms seem to be
superfluous ones.
But they are necessary for removing the ultraviolet
divergences at
the {\it following} steps of the iteration scheme (see
\cite{Ro1}--\cite{Ro3}
for more detailed discussion of the renormalization
in the given
iteration scheme). Since in the present work we
limit ourselves to
studying the first-step equations, we shall not write the
counterterms
and  terms of $P_1$ which are linear in sources,
as well as
the connections among them,
because they are necessary only at the following
steps of the
calculations.

Let us discuss leading approximation (5) in more detail.
The role
of the leading approximation is reduced to the
definition of
structure of the ground state --- the physical
vacuum of the theory.
Firstly, notice that eq.(6) possesses a variety of
solutions
$\{v\}$,
 and each of them defines a solution
of eq.(4), and, correspondingly, iterative solution
of SDE (3).
The trivial solution of the characteristic equation
$v\equiv~0$
(i.e., $G_0=~1$) leads, nevertheless, to the non-trivial
solution of SDE.
In this case the first-step Green functions  are
$\Delta = (m^2 + \partial^2)^{-1} = \Delta_c, \; d\equiv
\Phi\equiv 0$,
and the considered iteration scheme defines the
reconstructed series of
the perturbation theory in coupling constant over the
trivial perturbative
vacuum.
Though  {\it a priori} one cannot exclude
coordinate-dependent solution
of eq.(6), for our purposes it is sufficient to limit
ourselves
to the class of solutions $v = \mbox{const}$.
Each coordinate-independed $v$
 defines the Poincar\'{e}-invariant theory. At  $m^2 < 0$
this class of solutions describes the phase with
spontaneously
broken $U(1)$-symmetry. Really, eq.(6) takes the
familiar form
$$
\lambda v^*v + m^2 = 0.
$$
In this case eqs. (9) and (10) have  the solutions (in the
momentum space)
$$
\Delta (p) = \frac{1}{2} (\frac{1}{2\lambda
v^*v-p^2}-\frac{1}{p^2}), \;
d(p) =  \frac{1}{2}\frac{v^*}{v} (\frac{1}{2\lambda
v^*v-p^2}+\frac{1}{p^2}).
$$
Linear combinations
$$
\Delta_H = \Delta + \frac{v}{v^*}d = \frac{1}{2\lambda v^*v - p^2},\;
\Delta_G = \Delta - \frac{v}{v^*}d = -\frac{1}{p^2}
$$
correspond to Higgs boson and Goldstone boson propagators.
All the picture completely coincides with the
traditional approach
based on the study of the effective potential for model (2)
in the leading quasiclassical approximation.
\section{Higgs model}
The principle of local gauge invariance  requires
a gauge field
$A_\mu$ to be introduced, and Lagrangian (2) for the
gauge theory
is changed to the Lagrangian of Abelian Higgs model
\begin{equation}
{\cal L} = (\partial_\mu + ieA_\mu)\phi^*
(\partial_\mu -ieA_\mu)\phi
- m^2\phi^*\phi - \frac{\lambda}{2}(\phi^*\phi)^2 -
\end{equation}
$$
-\frac{1}{4}(\partial_\mu A_\nu  - \partial_\nu A_\mu)^2 -
\frac{1}{2\alpha}(\partial_\mu A_\mu)^2.
$$
In the theory with Lagrangian (12)
the system of SDEs for the generating functional
$G(j,J_\mu)$
 is of the form
\begin{equation}
(g_{\mu\nu}\partial^2 - \partial_\mu\partial_\nu
+ \frac{1}{\alpha}\partial_\mu\partial_\nu)
\frac{1}{i}\frac{\delta G}{\delta J_\nu} +
\end{equation}
$$
+ie\frac{\delta}{\delta j^*}(\partial_\mu +
e\frac{\delta}{\delta J_\mu})
\frac{\delta G}{\delta j}
-ie\frac{\delta}{\delta j}(\partial_\mu -
e\frac{\delta}{\delta J_\mu})
\frac{\delta G}{\delta j^*} + J_\mu G = 0,
$$
\begin{equation}
\frac{\lambda}{i}\frac{\delta^3 G}{\delta j^* \delta j \delta j^*}
- (m^2 + (\partial_\mu - e\frac{\delta}{\delta J_\mu})^2)
\frac{1}{i} \frac{\delta G}{\delta j^*}
+ jG = 0.
\end{equation}
Here $J_\mu(x)$ is the source of field $A_\mu(x)$.
Let us apply the method of the preceding Section to the system of
eqs.(13)-(14)  and choose as a leading approximation a
system of equations
with "constant coefficients"
$$
(g_{\mu\nu}\partial^2 - \partial_\mu\partial_\nu
+ \frac{1}{\alpha}\partial_\mu\partial_\nu)
\frac{1}{i}\frac{\delta G_0}{\delta J_\nu} +
$$
$$
+ie\frac{\delta}{\delta j^*}(\partial_\mu +
e\frac{\delta}{\delta J_\mu})
\frac{\delta G_0}{\delta j}
-ie\frac{\delta}{\delta j}(\partial_\mu -
e\frac{\delta}{\delta J_\mu})
\frac{\delta G_0}{\delta j^*}=0,
$$
$$
\frac{\lambda}{i}\frac{\delta^3 G_0}
{\delta j^* \delta j \delta j^*}
- (m^2 + (\partial_\mu - e\frac{\delta}{\delta J_\mu})^2)
\frac{1}{i} \frac{\delta G_0}{\delta j^*}
 = 0.
$$
A solution of this system is the functional
$$
G_0(j,J_\mu) = \exp i\int dx[V_\mu(x)J_\mu(x) +
v^*(x)j(x) +
j^*(x)v(x)],
$$
where $V_\mu$ and $v$ are solutions of the system of characteristic
equations
$$
(g_{\mu\nu}\partial^2 - \partial_\mu\partial_\nu
+ \frac{1}{\alpha}\partial_\mu\partial_\nu)V_\nu
-iev(\partial_\mu + ieV_\mu)v^* +
iev^*(\partial_\mu - ieV_\mu)v
= 0,
$$
$$
\lambda v^*v^2 + (m^2 + \partial^2 -
2ieV_\mu\partial_\mu
- ie\partial_\mu V_\mu - e^2 V^2)v = 0.
$$
Below we shall consider only the class of solutions
$v~=\mbox{const}\neq~0$.
If, in addition, one imposes the subsidiary condition
$\partial_\mu V_\mu = 0$, the system of
characteristic equations
has the form
\begin{equation}
(\mu^2 + \partial^2) V_\mu = 0,
\label{CharV}
\end{equation}
\begin{equation}
\lambda v^*v + m^2 - e^2V^2 = 0,
\label{Charv}
\end{equation}
where $\mu^2 = 2e^2v^*v$.
A system of iteration scheme equations is
$$
(g_{\mu\nu}\partial^2 - \partial_\mu\partial_\nu
+ \frac{1}{\alpha}\partial_\mu\partial_\nu)
\frac{1}{i}\frac{\delta G_n}{\delta J_\nu}
+ie\frac{\delta}{\delta j^*}(\partial_\mu +
e\frac{\delta}{\delta J_\mu})
\frac{\delta G_n}{\delta j} -
$$
$$
-ie\frac{\delta}{\delta j}(\partial_\mu -
e\frac{\delta}{\delta J_\mu})
\frac{\delta G_n}{\delta j^*}
= - J_\mu G_{n-1} + \mbox{counterterms},
$$
$$
\frac{\lambda}{i}\frac{\delta^3 G_n}
{\delta j^* \delta j \delta j^*}
- (m^2 + (\partial_\mu - e\frac{\delta}{\delta J_\mu})^2)
\frac{1}{i} \frac{\delta G_n}{\delta j^*}
 = -jG_{n-1} + \mbox{counterterms}.
$$
A solution of the first-step equations $(n=1)$ will be
looked for the form
$$
G_1 = P_1 G_0,
$$
where $P_1$ is a second-order polynomial
in sources $j$ and $J_\mu$:
$$
P_1(j,J_\mu) =
\int dx dy \bigl\{ij^*(x)\Delta(x,y)j(y) +
\frac{i}{2}\bigl(j(x)d(x,y)j(y)+
$$
\begin{equation}
 + j^*(x)\bar d(x,y)j^*(y)\bigr)
+\frac{1}{2i} J_\mu(x)D_{\mu\nu}(x,y)J_\nu(y) +
\end{equation}
$$
+i\bigl(J_\mu(x)B_\mu(x\mid y)j(y) + J_\mu(x)\bar
B_\mu(x\mid y)j^*(y)\bigr)
\bigr\}
+ \mbox{linear terms}.
$$
With characteristic eqs. (15)-(16) taken into account, the
first-step equations lead to the system of equations
for the
coefficient functions of polynomial (17):
\begin{equation}
L_{\mu\nu}D_{\nu\lambda} +
ie(v\partial_\mu B^T_\lambda -
v^*\partial_\mu \bar B^T_\lambda) -
2e^2 V_\mu(vB^T_\lambda +
v^*\bar B^T_\lambda) = g_{\mu\lambda},
\end{equation}
\begin{equation}
L\bar B^T_\nu + \lambda v^2 B^T_\nu + v(ie\partial_\mu +
2e^2V_\mu)D_{\mu\nu} = 0,
\end{equation}
\begin{equation}
L_{\mu\nu}B_\nu + ie\partial_\mu(v^*\Delta - vd) +
2e^2V_\mu(v^*\Delta + vd) = 0,
\end{equation}
\begin{equation}
L_{\mu\nu}\bar B_\nu +
ie\partial_\mu(v^*\bar d - v\Delta^T) +
2e^2V_\mu(v^*\bar d + v\Delta^T) = 0,
\end{equation}
\begin{equation}
L\Delta + \lambda v^2d - v(ie\partial_\mu +
2e^2V_\mu)B_\mu = 1,
\end{equation}
\begin{equation}
L\bar d + \lambda v^2\Delta^T - v(ie\partial_\mu +
2e^2V_\mu)\bar B_\mu = 0.
\end{equation}
Here
$L_{\mu\nu}\equiv(\mu^2+\partial^2)g_{\mu\nu}~
-\partial_\mu\partial_\nu+
\frac{1}{\alpha}\partial_\mu\partial_\nu$ and
$L\equiv\partial^2+\lambda~v^*v-2ieV_\mu\partial_\mu.$
The upper index $T$ means transposition:
$\Delta^T(x,y)\equiv~\Delta(y,x)$ and so on.
Besides  eqs.(18)-(23), three conjugated equations exist
which follow from
the Schwinger-Dyson equation
$$
\frac{\lambda}{i}\frac{\delta^3 G}
{\delta j\delta j^* \delta j}
- (m^2 + (\partial_\mu + e\frac{\delta}{\delta J_\mu})^2)
\frac{1}{i} \frac{\delta G}{\delta j}
+ j^*G = 0,
$$
which is conjugated to eq.(14). These equations differ from
eqs.(19),
(22) and (23) in the substitution
$$
L\rightarrow L^*,\;B_\lambda\leftrightarrow\bar B_\lambda,
\;\Delta\leftrightarrow\Delta^T,\;
d\leftrightarrow\bar d,\;v\leftrightarrow v^*,\;
i\rightarrow -i.
$$
For the investigation of the system of equations
for coefficient
functions it is useful to introduce the linear combinations
$$
C^\pm_\lambda = vB^T_\lambda \pm v^*\bar B^T_\lambda,\;
\Delta_H = \Delta + \frac{v}{v^*}d,\;\Delta_G =
\Delta -
\frac{v}{v^*}d.
$$
First of all, notice, that with the formula
$$
\partial_\mu~L_{\mu\nu}=
(\mu^2+\frac{1}{\alpha}\partial^2)\partial_\nu
$$
one can easy calculate the longitudinal part of gauge
field propagator $\partial_\mu~D_{\mu\nu}$ from eqs.
(18), (19)
 and from an equation which is conjugated to eq.(19).
 The result is
$$
\partial_\mu D_{\mu\nu} = \alpha\frac{\partial_\nu}
{\partial^2}.
$$
Therefore, the longitudinal part of gauge field
propagator is not
renormalized. Below we shall work in the transverse gauge
$\alpha~=~0.$ At $\alpha\rightarrow~0$
$$
L_{\mu\nu}D_{\nu\lambda} = (\mu^2+\partial^2)
D_{\mu\lambda}
+ \partial_\mu\partial_\lambda /\partial^2,
$$
$$
\partial_\mu C^\pm_\mu = 0,
$$
and the system of equations is simplified. Excluding
the combination
 $C^+_\lambda$ from the system of
eqs. (18), (19) and from one conjugated  to (19),
we obtain the system of equations
\begin{equation}
(\mu^2+\partial^2)D_{\mu\nu} + 4e^2\mu^2V_\mu\Delta_1\star
(V_\rho D_{\rho\nu})
 + ie\bigl[\partial_\mu C^-_\nu +
4e^2 V_\mu\Delta_1\star(V_\rho\partial_\rho C^-_\nu)\bigr]
=
\pi_{\mu\nu},
\end{equation}
\begin{equation}
\partial^2 C^-_\nu +
4e^2 V_\mu\partial_\mu\Delta_1\star
(V_\rho\partial_\rho C^-_\nu)
= 4ie\mu^2 V_\mu\partial_\mu\Delta_1\star
(V_\rho D_{\rho\nu}).
\end{equation}
Here $\Delta_1\equiv~(\partial^2+2\lambda~v^*v)^{-1}$
and $\pi_{\mu\nu}\equiv~g_{\mu\nu}-
\partial_\mu\partial_\nu/\partial^2.$
The symbol $\star$ denotes the operator multiplication:
$(a\star~b)(x,y)\equiv\int~dz~a(x,z)b(z,y).$
Excluding $B_\mu$ and $\Delta_G$ from eqs.(20), (22) and
from that conjugated
to (23),  one obtains (after simple transformations)
the equation for Higgs propagator
\begin{eqnarray}
(\partial^2 + 2\lambda v^*v)\Delta_H +
4e^2 V_\mu\partial_\mu\Delta_0\star
(V_\nu\partial_\nu\Delta_H) +
 \nonumber\\
 +4e^2\mu^2 V_\mu D^c_{\mu\nu}\star(V_\nu\Delta_H)
= 1 + 2ieV_\mu\partial_\mu\Delta_0.
\end{eqnarray}
Here
 $D^c_{\mu\nu}
 \equiv~(\mu^2+\partial^2)^{-1}\star\pi_{\mu\nu}$
and $\Delta_0\equiv~1/\partial^2.$
At $V_\mu\equiv 0$ eqs.(24)-(26) have the solution
$$
D_{\mu\nu} = (\mu^2+\partial^2)^{-1}\star\pi_{\mu\nu},\;
 \Delta = (\partial^2+2\lambda~v^*v)^{-1},\;
C^-_\nu = 0,
$$
which exactly corresponds to the result of Higgs model
in transverse
gauge for usual approach when the effective potential
is used.
Let us consider now a generalization of the Higgs mechanism
taking into account the solutions with
$V_\mu\not\equiv~0$.
Firstly, notice, that such solutions exist.
Indeed, besides the solution
$V_\mu=0,\;\lambda(v^*v)=-m^2$,
the characteristic equations (15)-(16)
have the class of solutions
\begin{equation}
V_\mu(x) = V^c_\mu\cos ax + V^s_\mu\sin ax,
\end{equation}
where constant vectors $V^c_\mu,\;V^s_\mu$ and
$a_\mu$ satisfy
the conditions
\begin{equation}
(V^c)^2 = (V^s)^2 = \frac{\lambda v^*v + m^2}{e^2},\;
a^2 = \mu^2,\;(V^cV^s) = (aV^c) = (aV^s) = 0.
\end{equation}
Each solution
  $V\equiv( V_\mu,v)$ of the characteristic equation system
  defines some partial iterative solution
  $G_V$ of the SDEs.
This solution will be referred to as  corresponding to a
partial mode
$\mid~V>$.
Obviously, the choice of a separate partial mode with
$V_\mu\neq~0$ as a leading approximation to the physical
vacuum ("a candidate for the physical vacuum") does not
ensure Poincar{\' e}-invariance of the theory. Notice,
however, that SDEs  are the linear functional-differential
equations for the generating functional, and any
superposition of partial solutions $\sum G_V$  is also a
solution of these equations. So we can choose a
superposition of partial modes as a candidate for the
physical vacuum, and choose the generating
functional of the
physical Green functions as the superposition
$$
<0\mid 0>_J = G(J) = \sum_{\{V\}} G_V(J),
$$
corresponding to some class
 $\{V\}$ of solutions of the characteristic equations.
We shall suppose  this superposition can be
chosen in such a way
that all contributions, breaking the
Poincar\'e-invariance,
are mutually
canceled, and the resulting  theory turns out to be
Poincar\' e-invariant.
For instance, the  expectation value of the gauge field
should disappear:
 \begin{equation}
<0\mid  A_\mu\mid 0> = \frac{1}{i}\frac{\delta G}{\delta
J_\mu}\Bigg\vert_{J=0}= \frac{1}{i}\sum_{\{V\}}\frac{\delta
G_V}{\delta J_\mu}\Bigg\vert_{J=0} = 0,
\label{veA}
\end{equation}
in spite of  the contributions of separate partial
modes in
this vacuum expectation can be different from  zero.
Further, the higher derivatives of the  physical generating
functional $G(J)$, defining many-point functions, must be
translational-invariant after switching-off the
sources, etc.
It is not difficult to  make  condition (\ref{veA})
hold true.
For this we notice that $- V_\mu$ is a solution of the
characteristic  equations
(\ref{CharV})-(\ref{Charv}) as well as $V_\mu$ is, so for
obeying (\ref{veA}) in the leading approximation it is
sufficient to take superposition
$G_V+G_{-V}\sim \cos{\int}dxJ_{\mu}V_\mu$. Note, that
simultaneously the vacuum expectations of all odd monomials
in $V_\mu$  also turn to zero:
$$
<0\mid~{V_\mu}_1\cdots{V_\mu}_{2n+1}\mid~0>=0.
$$
As it easy to see,
there is also no problem with  condition (\ref{veA})  in the
higher orders.
The requirements for many-point functions are less trivial.
For instance, we must require
\begin{equation}
  <0\mid V_\mu(x)V_\nu(y)\mid 0>=f_{\mu\nu}(x-y).
\label{veV2}
\end{equation}
It is clear to make formula (\ref{veV2}) hold true the
required operation $\sum_{\{V\}}$ should be continual, i.e.,
should correspond to some integration.
 But for the calculation of the function
 $f_{\mu\nu}$ itself
 there is no necessity to specify this operation.
Really, due to subsidiary condition
$\partial_\mu~V_\mu=~0$
and characteristic eqs.(15)-(16) this function has the form
\begin{equation}
f_{\mu\nu} = \frac{\lambda v^*v + m^2}{3e^2}
\pi_{\mu\nu}\star f,
\label{fmunu}
\end{equation}
where the scalar function $f(x-y)$ is a
solution of the equation
$$
(\mu^2 + \partial^2)f = 0
$$
with initial condition $f(0)=1$, i.e.
\begin{equation}
f(x) = \frac{2}{\sqrt{\mu^2x^2}}J_1\bigl
(\sqrt{\mu^2x^2}\bigr),
\label{f(x)}
\end{equation}
where $J_1(z)$ is the Bessel function.
The same result for $f_{\mu\nu}$ can be obtained
by means of
the direct application of the averaging procedure to
solution (27) (see \cite{Ro4}).
The calculation of the expectation values of the
higher even
monomials in $V_\mu$ can be performed in similar manner.
  Physical  propagators must be built by means of the same
  operations of partial mode superposition:
$$
\Delta_H(x-y) = \sum_{\{V\}}\Delta_H(x,y\mid V),
$$
where $\Delta_H(x,y\mid V)$ is a solution of eq.(26) for
some partial solution $V$ of the characteristic equations,
and so on.
Full solving of eqs.(24)-(26) with consequent transition to
the physical vacuum presents  a difficult problem since at
$V_\mu\not\equiv~0$ this system of equations  is the
complicated system of integro-differential equations.
For its approximate solving  we change the product
$V_\mu V_\nu$ in eqs.(24)-(26)
by its expectation value on the physical vacuum:
$$
V_\mu(x)V_\nu(y)\Rightarrow <0\mid V_\mu(x)V_\nu(y)\mid 0>=
f_{\mu\nu}(x-y),
$$
i.e., we shall use some type of  mean-field
approximation. Though this approximation can be shown rather
crude, nevertheless such type of approximations   describe
properly many statistical systems
and with extraordinary success it works in microscopical
theory of superconductivity \cite{Bog}, \cite{Tin}.
The physical vacuum of the Higgs model is a relativistic
superconductor,
and it is reasonable to suppose such approximation will be
not  bad in the given case as well.
The equation for the Higgs propagator $\Delta_H$
in this approximation has the form
\begin{equation}
(\partial^2 + 2\lambda v^*v)\Delta_H +
4e^2 (f_{\mu\nu}\partial_\mu\partial_\nu\Delta_0)\star
\Delta_H
+ 4e^2\mu^2 (f_{\mu\nu} D^c_{\mu\nu})\star\Delta_H = 1.
\label{eqH}
\end{equation}
 Since
$\partial_\mu f_{\mu\nu}=~0,$ it follows from eq.(25)
that in our approximation $C^-_\nu=~0,$
and the equation for the gauge field propagator takes
the form
\begin{equation}
(\mu^2+\partial^2)D_{\mu\nu} +
4e^2\mu^2(f_{\mu\rho}\Delta_1)\star
 D_{\rho\nu} = \pi_{\mu\nu}.
\label{eqD}
\end{equation}

Equations (\ref{eqH}) and (\ref{eqD}) are
translational-invariant and are easily
solved in the momentum space:
\begin{equation}
\tilde{\Delta}_H(p)=[2{\lambda}v^*v-p^2+
\tilde \Sigma(p)]^{-1},
\;\tilde{D}_{\mu\nu}(p)=[\mu^2-p^2+
\tilde \Pi(p)]^{-1}(g_{\mu\nu}-
\frac{p_\mu p_\nu}{p^2}),
\end{equation}
where
$$
\Sigma = 4e^2[(f_{\mu\nu}\partial_\mu\partial_\nu\Delta_0)+
\mu^2(f_{\mu\nu}D^c_{\mu\nu})],\;
\Pi = \frac{4}{3}e^2\mu^2(f_{\mu\nu}\Delta_1)\star
\pi_{\mu\nu}.
$$
Further calculation of the propagators in the
momentum space
is reduced
to the straightforward calculation of "single-loop
integrals" like\\
$\int dq\tilde f_{\mu\nu}(p-q)\tilde D_{\mu\nu}^c(q)$.
A location of propagator poles defines masses of particles.
A distinctive feature of the generalized Higgs
mechanism in
comparison
with the usual one is the possibility to model the
triviality of
 $\phi^4_4$-theory, i.e., we can tend
 $\lambda$ to zero but the masses of Higgs and gauge bosons
will retain non-zero values.
At $\lambda\rightarrow~0$ the admissible values of
the parameter $m^2$ lie in the region
$$
-\infty < m^2 \le -6\mu^2.
$$
(At other values of the parameter
 $m^2$ the particle masses became complex-valued.)
When $m^2$ lies in the above region, the Higgs boson mass
$m_H$
and the gauge boson $m_A$ vary in the limits
$$
4.7\mu^2  > m^2_H \ge 4\mu^2,
$$
$$
\infty > m^2_A \ge 3.04\mu^2,
$$
and we obtain the triviality bound
\begin{equation}
m_H \le 1.15 m_A.
\label{bound}
\end{equation}
\section{Conclusion}
In the present work we have considered a generalization of
the Higgs mechanism
which takes into account the vacuum configurations with
non-zero expectation values of vector field. A contribution
of such configurations
 gives a possibility to keep a principal
physical result of the Higgs model --- a generation of gauge
field mass --- in spite of the triviality of
$\phi^4_4$-interaction.
This generalized Higgs mechanism imposes on
 the Higgs mass the very strict triviality
bound (\ref{bound}). A distinctive feature of the proposed
approach is the fact
that the triviality bound exists even for the simple Abelian
Higgs model
and does not require, contrary to the usual approach
  \cite{Callaway}, \cite{Triv},
 the scalar sector to be treated as an effective
 approximation
for some Grand unified theory.
The proposed generalization of the Higgs mechanism is
also possible,
of course,  for a non-Abelian theory. A main complication
for a non-Abelian theory is a drastic growing of the
"non-linearity degree"
of the characteristic equations. Nevertheless,   the strict
triviality
bound similar to (\ref{bound}) may remain for a non-Abelian
theory too.
At the same time, taking into account the vector
vacuum configuration
for a non-Abelian theory we acquire a principally
new possibility
of avoiding the consideration of the scalar sector at all:
a generation of gauge field mass
can be made dynamically. A consideration of such
models became quite
actual in connection with unsuccessful searches
of the Higgs boson
and leads to non-trivial phenomenological consequences
(see, for example, \cite{Arb}).\\
$$
\ast\ast\ast
$$

Author is grateful to P.A.~Saponov for useful comments.
The work is supported in part by RFBR,
grant No.98-02-16690.

\end{document}